\documentclass[letterpaper,12pt]{article}

\usepackage{graphicx}
\usepackage[latin1]{inputenc}
\usepackage{amssymb}
\usepackage{color}
\usepackage{times}

\usepackage{natbib}
\bibpunct{}{}{,}{s}{,}{,}

\title{Using sociometers to quantify\\ social interaction patterns}

\author{Jukka-Pekka Onnela$^{1*}$, Benjamin N. Waber$^{2,3}$,\\
Alex (Sandy) Pentland$^{3}$, Sebastian Schnorf$^{2}$ \& David Lazer$^{4,2}$}
\date{}

\begin{document} 

\maketitle 

\vspace{-10mm}

\begin{center}
\begin{small}
\noindent 
$^1$ Department of Biostatistics, Harvard School of Public Health, Boston, MA 02115\\
$^2$Harvard Kennedy School, Harvard University, Cambridge, MA 02138, USA\\
$^3$MIT Media Laboratory, Cambridge, MA 02139, USA\\
$^4$Northeastern University, Boston, MA 02115\\
\end{small}
\end{center}

\baselineskip24pt

\textbf{Research on human social interactions has traditionally relied on self-reports. Despite their widespread use, self-reported accounts of behaviour are prone to biases and necessarily reduce the range of behaviours, and the number of subjects, that may be studied simultaneously. The development of ever smaller sensors makes it possible to study group-level human behaviour in naturalistic settings outside research laboratories.  We used such sensors, sociometers, to examine gender, talkativeness and interaction style in two different contexts.  Here, we find that in the collaborative context, women were much more likely to be physically proximate to other women and were also significantly more talkative than men, especially in small groups. In contrast, there were no gender-based differences in the non-collaborative setting. Our results highlight the importance of objective measurement in the study of human behaviour, here enabling us to discern context specific, gender-based differences in interaction style.}

Research on human social interactions has relied on observations reported by humans, and both self-reported data and observer-recorded data, with varying degrees of observer involvement, have been used to quantify interactions. A popular method in social psychology has been to count behaviours and code them with respect to various criteria \cite{forsyth}. Studies by Bales on small-group interaction, known as interaction process analysis, are a classic example dating back to the 1950's \cite{bales1,bales2}. It is now possible to actively instrument human behaviour to collect detailed data on various dimensions of social interaction \cite{olguin,eagle,vesp},  the removal of the human observer arguably resulting in a less invasive approach to the study of social behavior.  This is important, because the presence of an external observer, typically the researcher, may heighten people's self-consciousness and concerns with appearing in socially desirable ways, which for some people could include acting in gender-typed ways \cite{deaux}. Therefore, gender differences may be more likely when researchers are present. Alternatively, social desirability may lead to the opposite effect; for example, men may act in a more affiliative manner in front of a researcher \cite{deaux}.

Analysing human behaviour based on electronically generated data has recently become popular.  Electronic sensors can be used to complement or replace human observers altogether, and while they may convey a slight sense of surveillance, this perception is likely reduced as sensors get smaller and smaller, and consequently less obtrusive.  Here, we used ``sociometers,'' which are wearable devices that use a high-frequency radio transmitter to gauge physical proximity to others, and a microphone to track speech, to collect detailed information on social interactions within particular contexts. Early explorations with sociometers have shown that even short, sliced signals can be powerful predictors for human communication  \cite{olguin,curhan,gips,eagle_som}.  We used the radio transmitter to infer whether and for how long any two participants were proximate to one another. The strength of the received signal was used to estimate the distance between sociometers, and we used a cut-off value of signal strength that corresponded to a physical distance of approximately 3 meters. The sociometer did not store raw audio data, but rather computed audio features that were used to infer speaking time, measured in seconds, for each participant. Finally, the signals from the built-in accelerometers, the third stream of data recorded, were used to ascertain that the participants wore the devices throughout the period of observation by monitoring the level of energy associated with their movement (see Methods).

Distinct from measurement accuracy, electronic instrumentation also enables researchers to study larger groups than possible with human observers. For any system with $N$ interacting (social) agents, the number of potential pairwise interactions increases to leading order as $N^2$. While research in the related field of social networks has classically relied on human-administered surveys and questionnaires, these approaches scale poorly precisely because of this large number of pairwise comparisons that need to be queried to construct large sociocentric networks.  Mobile phone communication data have recently enabled the exploration \cite{onnela2007} and modelling \cite{kumpula2007} of large-scale social networks, and they have also been used to investigate sex differences in the age and sex composition of conversation partners \cite{palchykov2012}. 

The relationship between gender and language is complex and subtle \cite{wood}. Some recent studies on talkativeness show little difference between men and women \cite{mehl,liberman}, but older literature in higher education settings suggests otherwise \cite{good1,good2}. A recent meta-analysis suggests that men are more talkative than women \cite{leaper1}, while earlier meta-analyses give mixed results \cite{hyde}. Some of the variability in results can potentially be explained by the use of different measures for talkativeness.  A narrative review \cite{james} reached a different conclusion about gender differences in talkativeness, deducing that most studies of adult conversation contradict the notion that women are more talkative than men. Conversely, gender differences in talkativeness appeared least likely during informal non-task-oriented contexts, suggesting that the activity structure or context might influence the direction and magnitude of gender differences in talkativeness.  More accurate instrumentation, resulting in higher quality data, may help resolve some of these puzzles, and it could facilitate the discovery of novel social dynamical phenomena. Further, understanding gender-based differences in interaction style could have implications for organizational effectiveness or policy interventions. If, for example, women have a proclivity to associate with other women, this could pose challenges for their promotion in organizations with predominantly male executives \cite{ibarra}.

\section*{Results}
To explore the relationship between gender and context, we collected data in two settings from subjects who had given their written consent to participate, one in higher education and the other in a workplace.  In each Setting, the subjects wore identical sociometers. Setting 1 encompassed the first day (12 hours) of an intense one-week long collaborative exercise at the end of the first year of a two-year Master's program at a public policy school in a private US university.  The students, who had previously earned their first degrees and were now pursuing subsequent professional degrees, were required to process a large quantity of sophisticated readings and lectures, within a week, into a memo with a policy recommendation.  Communication with other students was allowed. The performance of the students in the exercise affected their final course grade. We collected and analysed data from 79 students (42 males, 37 females). Setting 2, in contrast to Setting 1, was entirely unstructured and non-collaborative. We collected data from 54 co-located employees (16 males and 38 females) at a call centre in a major US banking firm. We analysed their behaviour during 12 one-hour lunch breaks, spanning several weeks, which the employees would typically spend in a cafeteria, in relatively small groups, in the same building. As in Setting 1, there was ample space available for individuals to interact with others in groups if they chose to do so.

 We chose the two settings for three reasons. First, we wanted to create a contrast between the two settings in terms of their collaborative nature. The students in Setting 1 were highly focused on their assignment, a major component of their professional degree, and they had an incentive to interact with one another during the one-week period to enhance their knowledge of various areas relevant to the assignment. In contrast, while talking with colleagues in the cafeteria in Setting 2 might be socially desirable, these subjects were taking a break from their work and therefore arguably in a different social mode. Second, we wanted each setting to contain an intermediate number of participants. This meant that the individuals could not conceivably interact as one large group but instead interacted in various groups of different compositions, yet the numbers were not too large such that the individuals could share the same physical environment. In other words, the surrounding space did not impose a cutoff on group size. Third, although the proportion of men and women in the two settings is different, each contained a sufficiently large number of persons of both sexes such that anyone with a preference for interacting with a person of either sex had the opportunity to do so. 

\begin{figure}[]
\begin{center}
\includegraphics[width=0.5\linewidth]{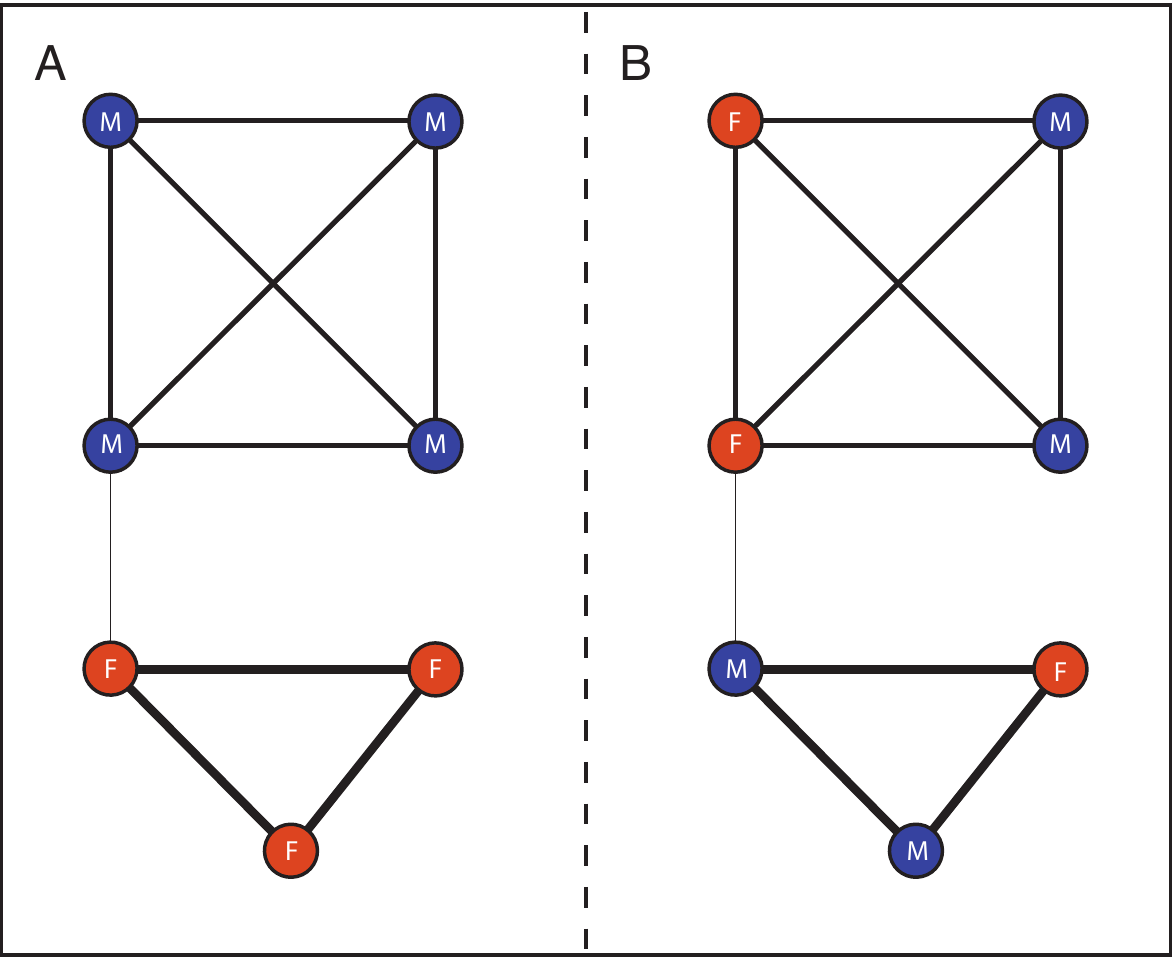}
\caption{Network construction from overlapping cliques and the gender-neutral null model. (A) The nodes represent individuals, where blue and red nodes correspond to males and females, also labelled with the letters M and F, respectively. The interaction network, at any instant, is made up of a number of small cliques, or fully connected subgraphs, which themselves are disjoint from one another. However, when examined over a time period exceeding the length of a single time window, these cliques may partially overlap. For example, the depicted network consists of a 4-clique (top) and a 3-clique (bottom), which are joined together by a 2-clique (a tie). Here tie strength is visualised by varying line widths (thicker lines correspond to longer  interactions). The number of male-male, female-male, and female-female ties in the observed (empirical) network in this schematic is $t_{\textrm{{\tiny MM}}} = 6$, $t_{\textrm{{\tiny FM}}} = 1$, and $t_{\textrm{{\tiny FF}}} = 3$. (B) A schematic of the null model. A realisation of the null model is obtained by keeping the structure of the network and its weights intact and randomly permuting the gender attributes of nodes. The number of male-male, female-male, and female-female ties in a realisation of the null model is denoted by $t_{\textrm{{\tiny MM}}}^{\ast}$, $t_{\textrm{{\tiny FM}}}^{\ast}$, and $t_{\textrm{{\tiny FF}}}^{\ast}$, respectively, and in this schematic $t_{\textrm{{\tiny MM}}}^{\ast} = 2$, $t_{\textrm{{\tiny FM}}}^{\ast} = 7$, and $t_{\textrm{{\tiny FF}}}^{\ast} = 1$. For investigating proximity, two variants of the null model were used, unconstrained and constrained; see Methods for details.} 
\label{fig:null}
\end{center}
\end{figure}

To carry out the analyses, in each Setting we first divided the 12 hours of data into 144 segments, corresponding to 5-minute time windows. In the resulting networks constructed from these data segments, any two individuals were linked if they had been proximate to one another for at least the duration of one full time window.  Encounters that did not fully cover at least one time window were deemed inconsequential and were not included as ties in the network.  In any network snapshot, constructed from proximity data over a single time window, the only structures present were (typically small) cliques, or fully connected subgraphs, which tied together the individuals who were in close physical proximity at that time. The cliques themselves, which comprise isolated nodes (1-clique) and isolated pairs (2-cliques) as special cases, were disconnected from one another. However, when examined over longer time periods consisting of multiple time windows, these cliques typically became connected as subsequent cliques bridged together nodes in antecedent cliques (see Fig.~\ref{fig:null}).  

\subsection*{Proximity}
Table \ref{tab:degree} tabulates the mean degrees measured in time windows for the subjects in Setting 1 and Setting 2, the degree of the subjects by their sex, as well as the degree of the subjects conditional on the sex of their interaction partners. We repeated our analyses using time windows of various widths and found the results to be remarkably robust. Using 5-minute windows, the mean degree of subjects was $4.5$ in Setting 1 and $7.1$ in Setting 2. In both settings, females and males had a similar number of connections, $4.8$ vs.~$4.1$ in Setting 1 and $7.1$ vs.~$7.2$ in Setting 2, respectively. The breakdown of degree by the sex of the conversation partners first looks different across the settings, but this can be explained by the different makeups of the two settings. In Setting 1, where there is approximately the same number of females and males (37 vs.~42), the mean degree of subjects to females and males is fairly similar ($2.5$ vs.~$2.0$). In Setting 2, the number of ties to females and males is very similar; the mean degree of subjects to females is $2.7$ times that to males ($5.2$ vs.~$1.9$), but these numbers are in agreement with the fact that in Setting 2 there are approximately $2.4$ times as many females as males (38 vs.~16). Based on these temporal averages, which ignore the duration or persistence of each pairwise interaction, females and males appear to behave similarly to one another, and they also appear to behave similarly across the two settings.

\begin{table}[htdp]
\caption{Mean momentary degree when the data are divided into window of length $T$ measured in minutes. Throughout this paper we have used $T=5$ minutes and the values of $T=2.5$ minutes and $T=10$ minutes are given for comparison. Each row in the table lists the mean degree for males (M), females (F), and everyone (M \& F) in Setting 1 and Setting 2. The columns M, F, and M \& F quantify how the degree gets split between male and female interaction partners. For example, in Setting 1 for $T=5$ minutes, the average male is connected to $4.1$ subjects, to $1.8$ males and $2.2$ females. Note that in Setting 1 there are 79 subjects (42 males, 37 females), whereas in Setting 2 there are 54 subjects (16 males and 38 females). Note that while node degrees are higher for longer time windows (higher $T$) as expected, the proportion of male and female ties remains almost unchanged.}
\begin{center}
\begin{tabular}{|l|c|c|c|c|c|c|c|c|c|}
\hline
\multicolumn{1}{|l|}{} & \multicolumn{3}{|c|}{Mean degree ($T=2.5$ min)} & \multicolumn{3}{|c|}{Mean degree ($T=5$ min)} & \multicolumn{3}{|c|}{Mean degree ($T=10$ min)}\\
\hline
 & M & F & M \& F  & M & F & M \& F  & M & F & M \& F \\
\hline
Setting 1 M & 1.4 & 1.7 & 3.1 & 1.8 & 2.2 & 4.1 & 2.3 & 2.9 & 5.2\\
Setting 1 F & 1.6 & 2.1 & 3.7  & 2.1 & 2.7 & 4.8 & 2.6 & 3.4 & 6.0\\
Setting 1 M \& F & 1.5 & 1.9 & 3.5 & 2.0 & 2.5 & 4.5 & 2.6 & 3.1 & 5.7\\
\hline
Setting 2 M & 1.1 & 4.0 & 5.1 & 1.5 & 5.7 & 7.2 & 2.0 & 7.7 & 9.6\\
Setting 2 F & 1.5 & 3.5 & 5.0 & 2.1 & 5.1 & 7.1 & 2.8 & 6.9 & 9.7\\
Setting 2 M \& F & 1.3 & 3.6 & 5.0  & 1.9 & 5.2 & 7.1 & 2.6 & 7.1 & 9.7\\
\hline
\end{tabular}
\end{center}
\label{tab:degree}
\end{table}

To incorporate the role of tie persistence in our analyses, we used the measured durations of proximity as tie strengths in the resulting aggregate network. We conjecture that whatever tendency there may be for the formation of MM, FM, and FF ties, the tendency should get stronger as we consider ties associated with longer physical proximity, i.e., as we move from potentially accidental short encounters to longer and arguably more deliberate encounters. We distinguished between male-male (MM), female-male (FM), and female-female (FF) ties, using $t_{\textrm{{\tiny MM}}}$, $t_{\textrm{{\tiny FM}}}$, and $t_{\textrm{{\tiny FF}}}$ to denote the count of each tie type present in a given window, respectively. To examine this hypothesis, we let $G(w)$ represent the overall aggregate proximity network where ties with $w_{ij} < w$ have been filtered out, leaving only stronger ties with $w_{ij} \ge w$ in place. 

\begin{figure}[]
\begin{center}
\includegraphics[width=0.75\linewidth]{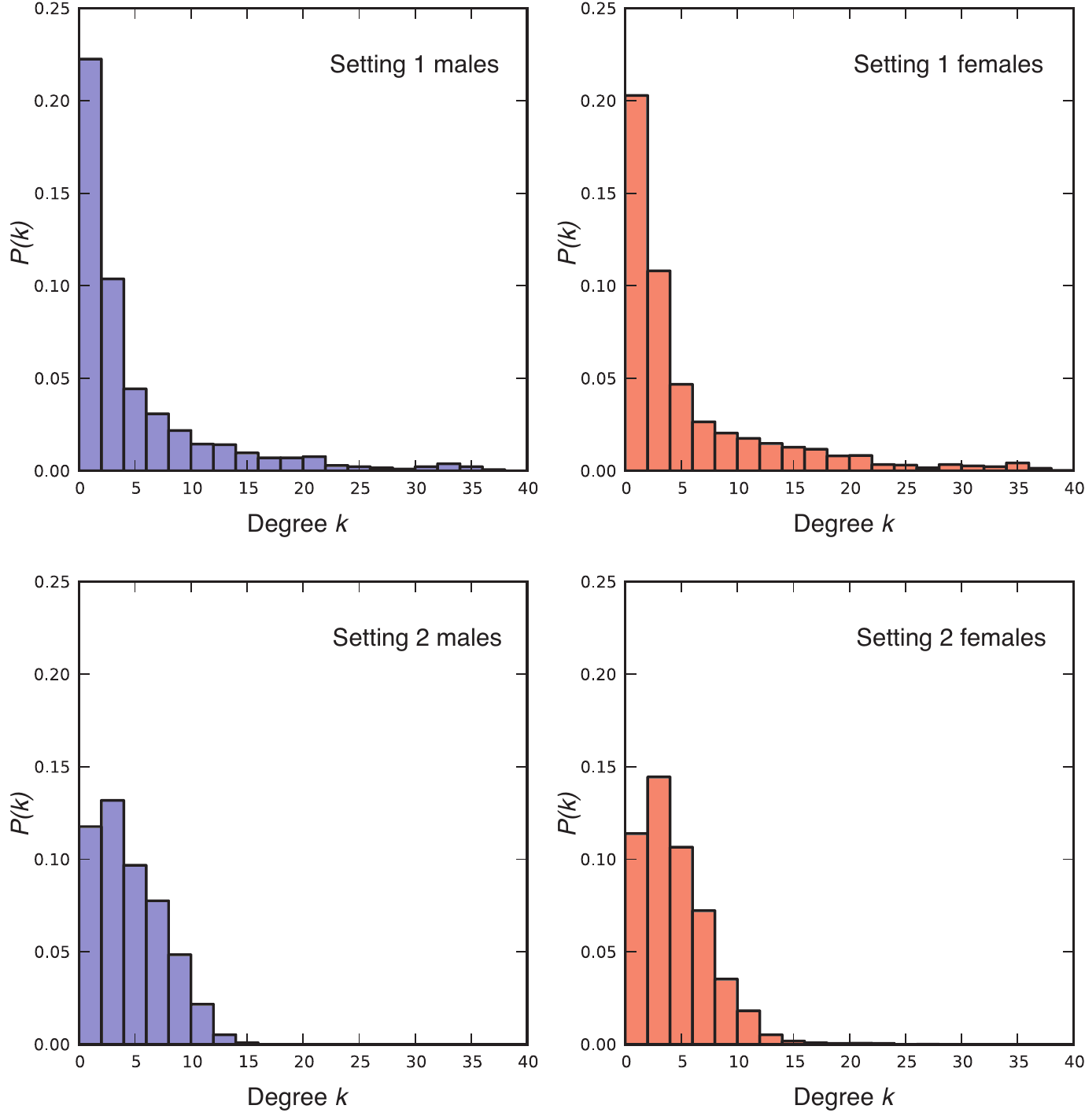}
\caption{ Distribution of momentary node degree (proximity) in each Setting for males and females using 5-minute window width. }
\label{fig:deg}
\end{center}
\end{figure}

The raw tie counts are however not very informative: (i) the total number of individuals in each setting is different; (ii) the number of males and females, and hence their proportion, is different across settings; and (iii) the counts are not adjusted for chance occurrence of ties, i.e., ties that would occur even in the absence of any gender-based preference. To address these issues, we defined a \emph{gender-neutral null model}: the tie counts, in each of the three categories and in each Setting, were normalised by dividing the observed tie counts by the expected tie counts generated under the null model, consisting of random permutations of the gender attributes (see Fig.~\ref{fig:null}).  We carried out two different variants of permutation: (i) the unconstrained permutation that is agnostic about possible differences in the degrees of men and women, and (ii) the constrained permutation that preserves the empirically observed mean degrees for men and women (see Methods). The unconstrained permutation implicitly assumes that node degree is independent of gender, and in this sense does not control for potential differences in degree between men and women. Fig.~\ref{fig:deg} shows the degree distributions to be very similar for men and women within each Setting, although they are quite different across the settings. However, since we reported some differences in mean degree between men and women above, we carried out proximity analyses using both unconstrained and constrained permutations. 

\begin{figure}[]
\begin{center}
\includegraphics[width=0.89\linewidth]{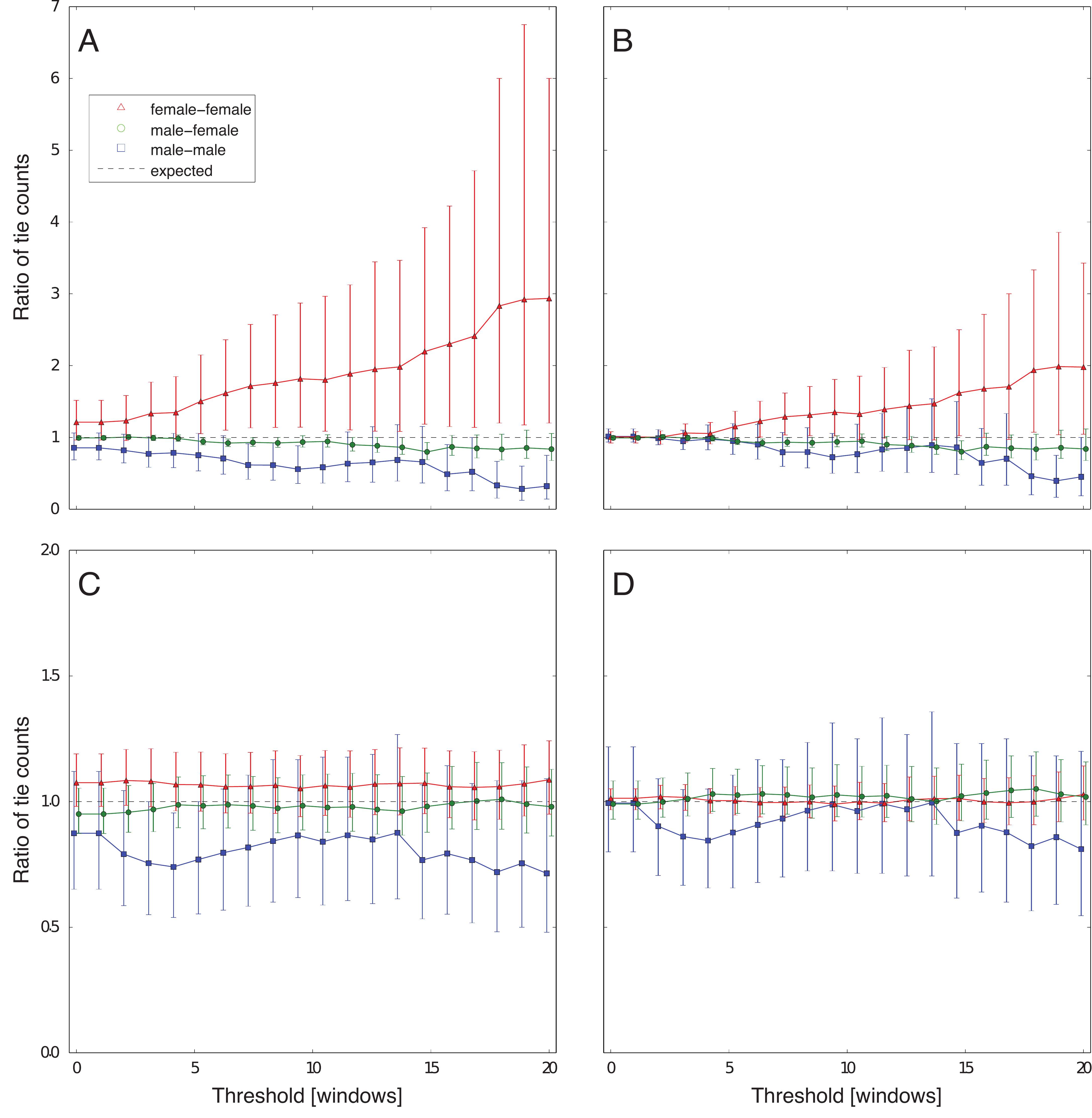}
\caption{Proportion of MM, FM, and FF ties as a function of threshold weight. Top row corresponds to Setting 1 and bottom row to Setting 2; left column corresponds to unconstrained permutation and right column to constrained permutation. The markers indicate the tie count ratios, obtained by dividing the number of observed tie counts by the number of expected tie counts, for female-female, female-male, and male-male ties, where the expected tie counts have been generated by the gender-neutral null model (see Materials) in Setting 1 (top row) and Setting 2 (bottom row). The dashed horizontal line indicates the value of tie count ratios expected under the null model ($r=1$). The vertical lines show the extent of variation in each category by displaying where 90\% of the tie count ratios fall under the null. The horizontal axis gives the threshold value $w$ in terms of proximity, such that ties with duration shorter than $w$ are excluded. In contrast to Setting 2, strong ties in Setting 1 tend to consist predominantly of two females. Note that the panels have different vertical scales across settings. The $x$-axis is measured in units of window width, in this case 5 minutes.}
\label{fig:prox}
\end{center}
\end{figure}

We show the resulting relative proportions of MM, FM, and FF ties in Fig.~\ref{fig:prox}, where we vary the value of the threshold from 1 to 20 window widths, i.e., from $5$ to $100$ minutes. While weak (short aggregate proximity) ties may be ``accidental,'' strong ties (extensive aggregate proximity) more likely are evidence of intended social interactions. We found that in Setting 1, there is a systematic over-representation of FF ties and an under-representation of MM ties. Further, this over-representation of FF ties increases monotonically with the threshold weight $w$ as the following results, based on 10,000 permutation replications, demonstrate. In the non-thresholded network (threshold $w=0$), there is only weak evidence to suggest that FF-ties might be over-represented: We obtain the ratio $1.21$ (90\% CI: 0.96, 1.52) under unconstrained permutation and $1.01$ (90\% CI: 0.92, 1.08) under constrained permutation. However, in the strongly thresholded network (threshold $w=20$), FF-ties are substantially over-represented: We obtain the ratio $2.94$ (90\% CI: 1.20, 6.00) under unconstrained permutation and $1.98$ (90\% CI: 1.04, 3.43) under constrained permutation. In contrast to Setting 1, in Setting 2 there is no perceptible statistically significant relationship between the frequency of FF, FM, or MM ties. In the non-thresholded network, we obtain the ratios $1.07$ (90\% CI: 0.98, 1.19) and $1.01$ (90\% CI: 0.97, 1.05) for unconstrained and constrained permutation, respectively; the corresponding numbers for strongly thresholded network are $1.09$ (90\% CI: 0.95, 1.24) and $1.03$ (90\% CI: 0.93, 1.14).

It is informative to examine these numbers in the context of mean degree. In Setting 1, for the non-thresholded network (threshold $w = 0$), the mean degrees of women and men are $45.0$ and $36.4$, respectively, the ratio of them being $1.23$; for the maximally thresholded network (threshold $w = 20$), the corresponding mean degrees are $4.2$ and $2.8$, resulting in a ratio of $1.49$. In Setting 2, for the non-thresholded network the mean degrees are $15.1$ and $13.4$, giving a ratio of $1.12$; for the maximally thresholded network, these numbers are $8.5$, $8.0$, yielding a ratio of $1.06$. Taken together, in Setting 1 women appear to have more high-persistence ties than men do, whereas in Setting 2 this is not the case.

\subsection*{Talkativeness}
\begin{figure}[]
\begin{center}
\includegraphics[width=0.6\linewidth]{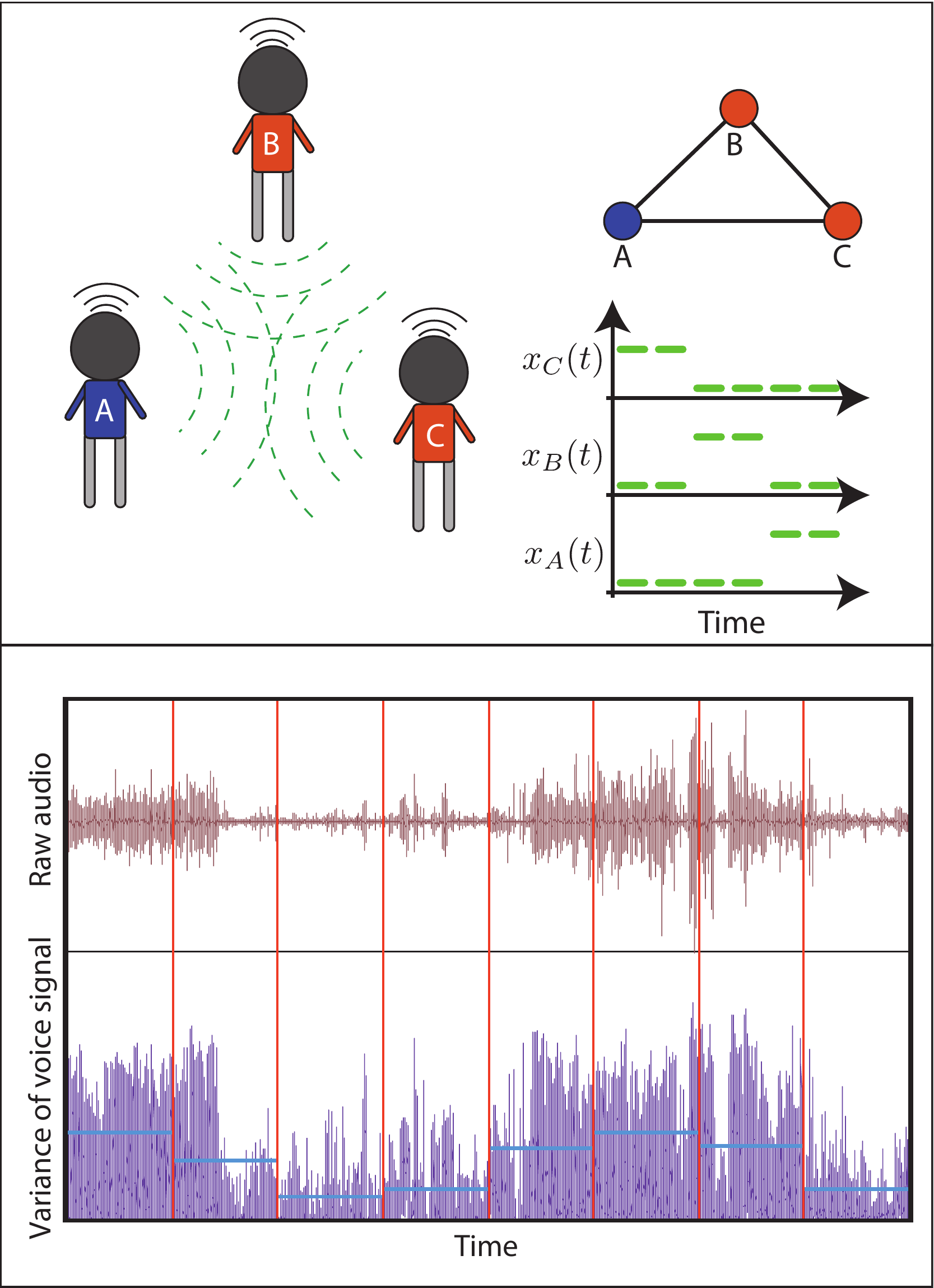}
\caption{Using sociometers to study physical proximity and talkativeness. Top: Every participant in each Setting was wearing an identical sociometer, which had a built-in radio transmitter and microphone. The radio transmitter was able to sense others who were within a radius of approximately 3 meters from the subject. In the corresponding network representation, a tie connects two individuals if they have been within this distance for at least one time window. The built-in microphones were used to assess the talkativeness of each individual, giving rise to the metric $x(t)$ for each individual as shown. Bottom: Schematic of the construction of the audio signals $x_i(t)$ and $y_i$ for node $i$. The raw audio signal, as recorded by the built-in microphone and temporarily stored on the sociometer for processing. This raw signal is first channeled to an array of band-pass filters encapsulating the range of typical human speech (see Methods), and the variance of the resulting human voice signal is shown. We then divide the temporal domain into time windows, and the average of the variance (lower panel) within each window gives the audio signal $x_i(t)$ in  time window $t$, pictorially represented by the horizontal lines within each time window. To express the overall talkativeness of a person, we take the temporal average of $x_i(t)$ over all time windows, giving rise to $y_i$ for subject $i$.}
\label{fig:pa}
\end{center}
\end{figure}

We then moved to examine talkativeness. The talkativeness of individuals is computed in a large number of time windows; although the raw data are collected at 750~Hz, the audio features are calculated at 50~Hz, still a high frequency. Instead of dealing with the raw audio signal, we use the variance of the audio signal in a range of frequencies typically associated with human voice (see Methods and Fig.~\ref{fig:pa}), which is a more robust way to distinguish between whether the signal comes from the person wearing the sociometer, or whether it corresponds to ambient noise (e.g., someone else talking).

\begin{figure}[]
\begin{center}
\includegraphics[width=0.5\linewidth]{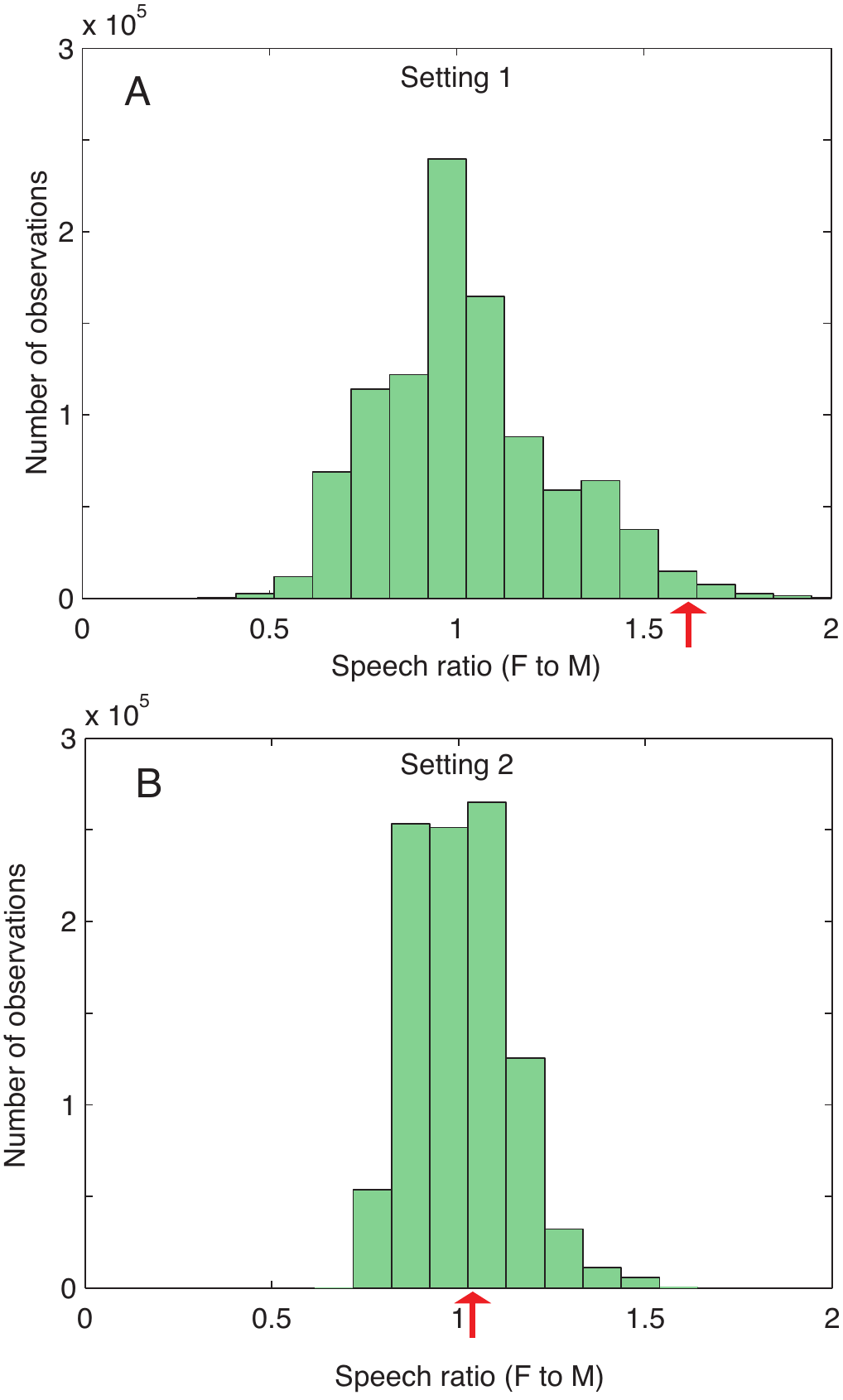}
\caption{Talkativeness by context. The observed speech ratios $r = \tilde{y}_F / \tilde{y}_M$, i.e., the median female talkativeness divided by the median male talkativeness, are indicated by the arrows at 1.62 and 1.04 for Setting 1 (A) and Setting 2 (B), respectively. The histograms show the expected distribution of speech ratios under the gender-neutral null model where we have permuted the gender attributes of nodes 1 million times (see Methods).}
\label{fig:talk}
\end{center}
\end{figure}

We quantify the talkativeness of a person by taking the average of the variance of voice signal in each window (indicated in Fig.~\ref{fig:pa} by the short horizontal lines in the lower panel), resulting in one data point per individual per window, denoted by $x_i(t)$, where $i=1,2,\ldots,N$ indexes the individual and $t=1,2,\ldots,T$ indexes the time window. The data from each setting can be represented as matrix $\mathbf{X}$, where the rows corresponds to the subjects ($i$) and the columns to the time windows ($t$). We express the average talkativeness of a person, male or female, by the temporal average $y_i = (1/T) \sum_{t=1}^T x_{it}$. To compare the talkativeness of males and females, we compute the median of the $y_i$ variables for males and females, resulting in $\tilde{y}_M$ and $\tilde{y}_F$, respectively. Computing the ratio $r = \tilde{y}_F / \tilde{y}_M$, i.e., the median female talkativeness divided the median male talkativeness, results in $r \approx 1.62$ for Setting 1 and $r \approx 1.04$ for Setting 2, which suggests that women are 62\% more talkative in the former context but only 4\% more talkative in the latter context. The reason for taking the median of the variable $y_i$ is that it is much less sensitive to outliers, such as exceptionally talkative individuals, than the mean, and hence better characterises a typical individual in a group. The distribution of the talk-ratio variable $r$ under the gender-neutral null model is shown in Fig.~\ref{fig:talk} (see also Methods). We found that the observed value of $1.62$ in Setting 1 is statistically significantly different from 1 ($p < 0.01$), whereas the value of $1.04$ in Setting 2 is not. We conclude that women are substantially more talkative in Setting 1, but there is no difference in the talkativeness of men and women in Setting 2.

\begin{figure}[h!]
\begin{center}
\includegraphics[width=0.5\linewidth]{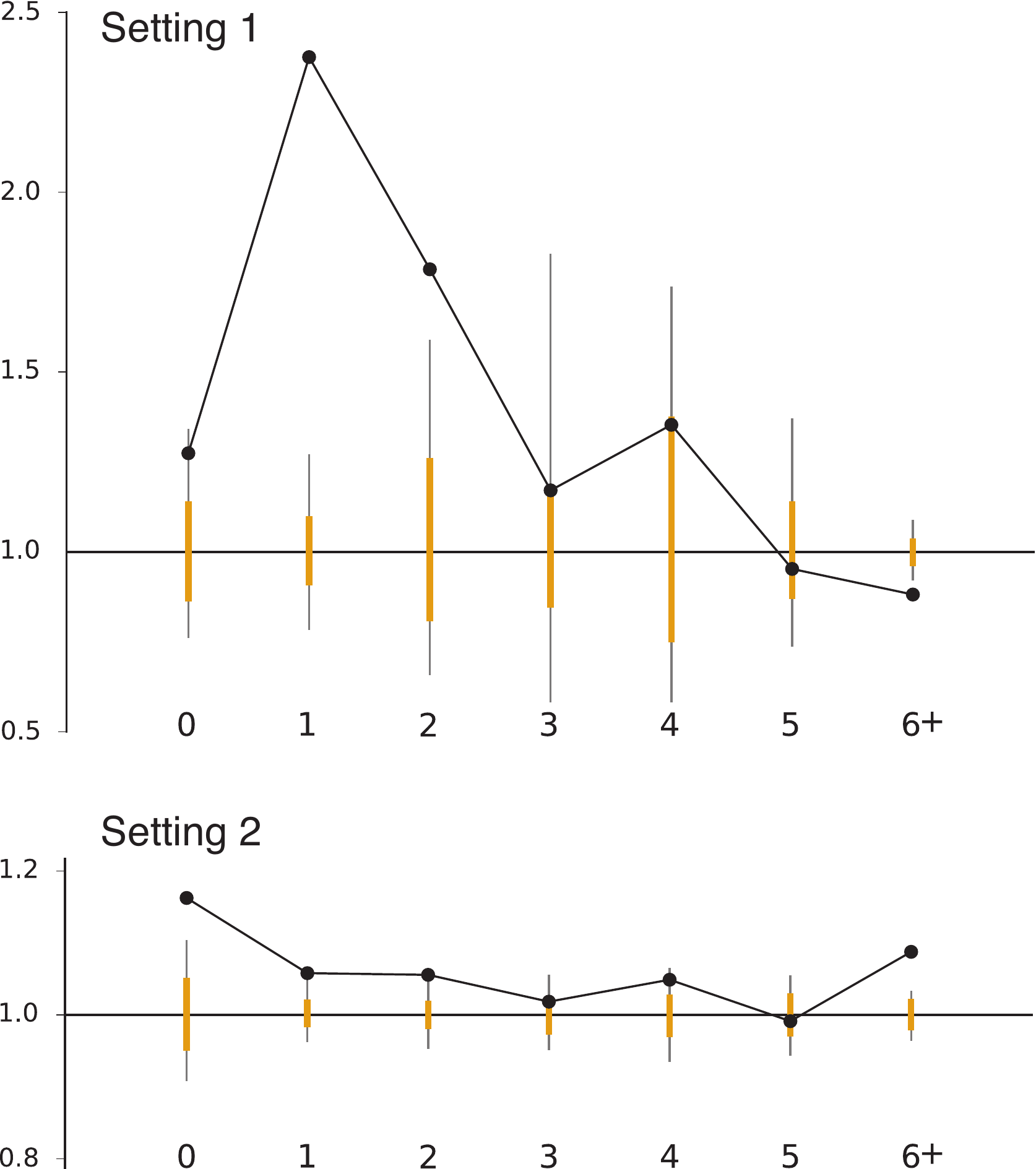}
\caption{Talkativeness ratio $r$ stratified by degree. For values of $r>1$ females talk more, whereas for values of $r < 1$ males talk more. The thick and thin horizontal lines show the simulated 50\% and 90\% confidence intervals, respectively, for the null model. Individuals with degree $k=0$ are not proximate to anyone in the study.}
\label{fig:degtalk}
\end{center}
\end{figure}

We combine the proximity and talkativeness data of sociometers to examine talkativeness as a function of interaction partners. Table \ref{tab:talk} and Figure \ref{fig:degtalk} give the values of the talkativeness ratio $r$ as a function of momentary degree, and also show the 50\% and 90\% confidence intervals. A large proportion of all interactions take place in small groups with just one or two interaction partners per subject.  In Setting 1, these small-group interactions ($k=1$ or $k=2$) make up 39\% of females' interactions and 46\% of males' interactions; in Setting 2, the small-group interactions are somewhat less prominent, making up 33\% of females' interactions and 28\% of males' interactions. (Note that for $k=0$, the subject is not interacting with anyone in the study but could be talking to someone outside the study, on the cell phone, etc.) Women talk significantly more than men in Setting 1, where $r_{k=1}=2.38$ and $r_{k=2}=1.79$, whereas in Setting 2 these differences are only slight ($r_{k=1}=1.06$ and also $r_{k=2}=1.06$). The difference in the talkativeness of women between Setting 1 and Setting 2 is therefore mostly associated with differences in the talkativeness of women in small-group settings. Furthermore, in Setting 1, the collaborative setting, there appears to be a decreasing trend in talkativeness as a function of group size. Women talk much more than men in small groups ($k=1$ and $k=2$) but less than men in large groups ($k=6+$). This trend is not present in Setting 2.

\begin{table}[htdp]
\caption{Talkativeness ratio $r$, as defined in the text, stratified by degree. The 90\% confidence intervals are given in square brackets. For values of $r>1$ females talk more, whereas for values of $r < 1$ males talk more. The fraction of individuals with the given degree $k$ is given by $p_k^M$ and $p_k^F$ for males and females, respectively. Note that any individual with degree $k$ is a member of a clique of order $k+1$. Individuals with degree $k=0$ are not proximate to anyone in the study. Statistically significant values have been marked with an asterisk ($^{\ast}$).}
\begin{center}
\begin{tabular}{|c|c|c|c|c|c|c|}
\hline
\multicolumn{1}{|c|}{} & \multicolumn{3}{|c|}{\textbf{Setting 1}} & \multicolumn{3}{|c|}{\textbf{Setting 2}} \\
\hline
Degree $k$ & $r_k$ & $p_k^M$ & $p_k^F$ & $r_k$ & $p_k^M$ & $p_k^F$ \\
\hline
0 & 1.27 [0.76, 1.34] & 0.12 & 0.14 & 1.16$^{*}$ [0.91, 1.10] & 0.08 & 0.06 \\
1 & 2.38$^{*}$ [0.79, 1.27] & 0.33 & 0.26 & 1.06 [0.96, 1.06] & 0.15 &  0.17 \\
2 & 1.79$^{*}$ [0.66, 1.59] & 0.13 & 0.13 & 1.06$^{*}$ [0.95, 1.04] & 0.13 &  0.16 \\
3 & 1.17 [0.58, 1.83] & 0.08 & 0.08 & 1.02 [0.95, 1.05] & 0.13 &  0.12 \\
4 & 1.35 [0.59, 1.73] & 0.05 & 0.06 & 1.05 [0.94, 1.06] & 0.09 &  0.12 \\
5 & 0.95 [0.74, 1.37] & 0.04 & 0.04 & 0.99 [0.94, 1.05] & 0.10 &  0.10 \\
6+ & 0.88$^{*}$ [0.92, 1.09] & 0.26 & 0.28 & 1.09$^{*}$ [0.97, 1.03] & 0.31 &  0.27 \\
\hline
\end{tabular}
\end{center}
\label{tab:talk}
\end{table}

Setting 1 also included a briefing meeting for the week's tasks, approximately 60 minutes long, which was omitted from the above analyses. This natural variation in the setting allowed us, even if only momentarily, to monitor the behaviour of the same set of individuals, the participants in Setting 1, in a modified large-group context. Notably, the talkativeness ratio fell from $r=1.62$ to $r=0.95$, i.e., men and women now appear equally talkative. This highlights the effect of switching the interaction context from a smaller group to a larger and more gender-mixed group, a finding that is compatible with the older literature on speaking and gender in education \cite{good1,good2}, and also consistent with our finding on men being more talkative in large groups ($k=6+$) in Setting 1.

\section*{Discussion}
The strongest effect discovered in our study is the difference in gender participation as a function of tie strength and group size. These results support a possible amendment to earlier findings of individual talkativeness, and suggest that future research on group performance should include these variables, if only to control for their effects.  Specifically, an earlier study, which did not consider the proximity of others, found no significant gender differences in individual talkativeness \cite{mehl}, compatible with our results for Setting 2. In Setting 1, women were much more likely to associate with other women than men, and thus were also more talkative (with the exception of the briefing). These findings are consistent with prior research indicating that women tend to have more interactive learning styles than men \cite{blum}. Our results also highlight the role of context. Constructionist and contextualist models of gender assert that activity is a highly influential moderator of gender-related variations in social behaviour \cite{deaux,leaper4,leaper5}. Our results clearly support the notion that gender differences in both physical proximity and talkativeness are strongly present in the more structured task-oriented context (Setting 1), whereas they completely disappear in unstructured non-task-oriented context (Setting 2).

Our results appear relevant also in a larger context. More and more problems are solved in groups, and recent studies have indeed shown that diverse groups of problem solvers, referring to groups of people with diverse tools and skills, consistently outperform groups of the best and the brightest, a finding captured by the aphorism ``diversity trumps ability'' \cite{page1,page2}. Research is also increasingly done in teams across nearly all fields, and teams typically produce more frequently cited research than individuals do, including the exceptionally high-impact research \cite{wuchty2007}. Another recent study suggests that a ``general collective intelligence factor'' of a group is not strongly correlated with the intelligence of the individual group members, but instead with the average social sensitivity of group members, the equality in distribution of conversational turn-taking, and the proportion of females in the group \cite{sandy}. In order for teams to maximise their diversity and hence performance, understanding the role of interaction context in the expression or suppression of gender-related diversity, in particular modes of communication, appears extremely important.

We have highlighted the collaborative vs. non-collaborative nature of the two settings because this was how we chose the two settings in the study. Given the observational nature of our study, we cannot however exclude other possible explanations for the observed differences in behavior. It is likely that the two environments differ along a number of dimensions, such as organizational culture, and it is also likely that the participants differ in their covariates, such as age, which is associated with gender-based homophily \cite{palchykov2012}. Future studies could collect and make use of a larger set of covariates on each participant, and then it might be possible to estimate causal effects by conditioning on those covariates \cite{morgan2007}.  Further, we anticipate that this type of instrumentation will allow the development of a corpus of datasets allowing the evaluation of an array of contextual variables (culture, organizational context, other task-based variables) that likely affect interaction patterns.

New technologies provide accurate and minimally invasive ways to instrument human behaviour, enabling the study of human interactions in more naturalistic settings outside research laboratories. The present study suggests a key contextual contingency \cite{james} in the interplay of gender and talkativeness. As our study is an exploration of the insights that novel instrumentation can provide, more research is needed to identify the underlying operative mechanisms. 

\section*{Methods}
\subsection*{Gender-neutral null model for proximity}
For each proximity network (Setting 1 and Setting 2), we randomly permuted the gender attributes 10,000 times starting from the non-thresholded (threshold $w=0$) network. We then proceeded to threshold each such network, for every value of the threshold $w$ keeping track of the resulting number of MM, FM, and FF ties, denoted by $t(w)_{\textrm{{\tiny MM}}}^{\ast}$, $t(w)_{\textrm{{\tiny FM}}}^{\ast}$, and $t(w)_{\textrm{{\tiny FF}}}^{\ast}$. Note that these are all functions of the threshold weight $w$. We then computed the ratio of the number of observed ties of a given type to the number of ties generated under the gender attribute permutation, resulting in 10,000 realisations of $r(w)_{\textrm{{\tiny MM}}}^{\ast} = t(w)_{\textrm{{\tiny MM}}} / t(w)_{\textrm{{\tiny MM}}}^{\ast}$, $r(w)_{\textrm{{\tiny FM}}}^{\ast} = t(w)_{\textrm{{\tiny FM}}} / t(w)_{\textrm{{\tiny FM}}}^{\ast}$, and $r(w)_{\textrm{{\tiny FF}}}^{\ast} = t(w)_{\textrm{{\tiny FF}}} / t(w)_{\textrm{{\tiny FF}}}^{\ast}$.

We carried out two different implementations of the null model. In the unconstrained permutation, every realization of the null model was accepted and used in subsequent computations. In contrast, in the constrained permutation, only those realizations of the permutation were accepted that resulted in average degrees for men and women that matched the corresponding empirical estimates within a pre-specified tolerance. To be clear, in the constrained permutation, both the realized average degree for men and the realized average degree for women had to match their empirical values. For the unconstrained permutation, the average degrees produced by the null model deviated up to 12\% form their empirical values, the precise numbers varying across men and women and across the two settings. For constrained permutation, we imposed a 2.5\% tolerance, in other words, the mean degree of men and the mean degree of women in the non-thresholded (threshold $w=0$) network had to be within $2.5$\% from the corresponding empirical values. In practice, for Setting 1 this means that average degrees deviate by approximately $\pm1$ from the empirical averages ($45.0$ for women and $36.4$ for men); in Setting 2, the deviations are about $\pm 1/3$ from the empirical averages ($15.1$ for women and $13.4$ for men).

\subsection*{Statistics on talkativeness}
To examine the statistical significance of talkativeness results, starting from the variables $y_i$, we performed a random permutation of the gender attributes of individuals, and then proceeded to calculate the speech ratio $r$ as described in the main text. For each such permutation, we obtain a ratio $r^{\ast}$, which is computed on the basis of the shuffled gender attributes. We repeat this process 1 million times, and the histograms of the resulting ratios are shown in Fig.~\ref{fig:talk}. In particular, we are interested in the 5th and 95th percentiles, which indicate the range of values of $r$ expected under the null model. For Setting 1, we obtain the range $[0.675, 1.471]$, and use the null distribution to obtain $p < 0.01$ for the observed ratio of $1.619$. In contrast, for Setting 2, the range is $[0.819, 1.235]$, such that the observed value $1.038$ falls squarely in the middle with $p=0.39$. Putting these results together, women are more talkative in Setting 1, whereas there is no decipherable difference in the talkativeness of men and women in Setting 2. We checked the robustness of our results to the length of the time window by dividing the data into $100, 200, \ldots, 1000$ windows, which had a negligible effect on the results.

The results reported above for Setting 1 all excluded the briefing, approximately 1 hour in duration. In order to study the talkativeness of individuals at the briefing, we excluded a small number of individuals who did not attend the event, since the students who were absent were not part of the same interaction context. We inferred briefing attendance by using the radio data component of the sociometers to construct six proximity networks, each based on a disjoint 10-minute time window at the time of the briefing. We then detected the largest connected component of each network which, given the range of the radio transmitters and the confines of the lecture room, gave us an accurate picture of who was present. We then repeated the above analysis on talkativeness, but included only those individuals who attended the briefing. This resulted in $r  \approx 0.95$, which suggests that there is no statistically significant difference in the talkativeness of men and women.

\subsection*{Sociometer accelerometer}
An accelerometer measures changes in experienced acceleration. The badge's 3-axis accelerometer signal is sampled at 50~Hz, which is able to capture a wide range of human movement, given that 99\% of the acceleration power during daily human activities is contained below 15~Hz \cite{mathie}. The range of values for the accelerometer signal varies between $-3g$ and $+3g$, where $g = 9.81$~m/s$^2$ is the gravitational acceleration. In our study, we used data from the accelerometer to determine if the participants wore the sociometers. Each accelerometer measured energy levels due to physical movement above the reference level of 1 unit \cite{olguin}, ascertaining that the subjects wore the sociometers as instructed.

\subsection*{Sociometer microphone}
The microphone within the badge did not store raw audio data, but rather computed audio features that were used to infer speaking patterns. The microphone in the sociometer was connected to an array of band-pass filters that divided the speech frequency spectrum into four octaves: (1) 85 to 222~Hz, (2) 222 to 583~Hz, (3) 583 to 1,527~Hz, and (4) 1,527 to 4,000~Hz. These frequencies encapsulate the range of typical human speech. In particular, in this study we examined the variation in the audio signal that arrived from sample to sample (the sampling frequency was 750~Hz). The more variation there is present in the signal, the more confident we can be that the associated signal is indeed human speech and not due to an external source. Audio features like these can be used not only to infer that a person is speaking, but they can also capture nonlinguistic social signals, such as interest and excitement \cite{pentland}.

\subsection*{Sociometer radio}
The built-in omni-directional 2.4~GHz radio was designed to detect physically proximate interactions. The radios sent a transmission once every minute that contained the ID of the sending badge, some synchronisation information, and error correction bytes. By measuring the received signal strength, it was possible to estimate the distance to the sender. We used a cut-off on the signal strength value to register people who were located within 3 meters of one another. This distance was deemed to be  appropriate for detecting a level of physical proximity that likely corresponds to an intentional social interaction. Since the subjects were free to move around the premises, depending on the given physical environment surrounding them, which would affect the transmission of radio waves, there is an error of 1.5 meters on the distance estimates \cite{sugano}. This means that we cannot rule out the possibility that two individuals at 4.5 meters would have appeared to be physically proximate and, similarly, it is possible that some individuals would have needed to be within 1.5 meters in order to have been registered as having been physically proximate. Due to these spurious detections, there are likely some false positive and some false negatives in our dataset. Identical instrumentation, i.e., the fact that each subject wore an identical device, ensures that there was no person-to-person variability in how distance (or any other behavioural signal) was measured.

\textbf{Acknowledgements:} We are grateful to K.~Ara, E.~S.~Bernstein, N.~A.~Christakis, N.~Katz, T.~Keegan, T.~Kim, A.~Mohan, D.~Olguin Olguin and S.~Place for their help at various stages. JPO is supported by NIH-1DP2MH103909-01 (Onnela).

\vspace{0.5cm}

\textbf{Author Contributions:} JPO and DL conceived the study; BNW and ASP programmed the sociometers and extracted the raw data; JPO performed data analysis; JPO, SS and DL wrote the manuscript. All authors read the manuscript and provided comments.

\vspace{0.5cm}

\textbf{Author Information:} Reprints and permissions information is available at www.nature.com/reprints. The authors declare no competing financial interests. Readers are welcome to comment on the online version of this article at www.nature.com/nature. Correspondence and requests for materials should be addressed to JPO (onnela@hsph.harvard.edu).

\end{document}